\documentclass[twocolumn,draft]{svjour3}
\usepackage{amsmath,amsfonts,amssymb,amscd,graphicx}

\newenvironment{Proof}{\par\noindent{\bf Proof.}}{\hfill $\Box$\par\smallskip}

\def\vett#1{\underline{#1}}
\def\e_#1{\vett e_#1}
\def\k_#1{\vett k_#1}

\def\Re{\mathbb R}      
\def\Real{\operatorname{Re}}
\def\Span{\operatorname{Span}}
\def\And{,\@\ldots\hskip-.4pt,\@}

\def\a{\alpha}
\def\b{\beta}
\def\g{\gamma}

\def\l{\lambda}

\def\vphi{\varphi}
\def\w{\omega}      
\def\x{\hat{\plus50 x\@}}

\def\hxtilde{\raise-1ex\hbox to0pt{$\scriptstyle\sim\hss$} \@\x{}}
\def\etatilde{\raise-1.4ex\hbox to0pt{$\scriptstyle\sim\hss$} \eta}
\def\utilde{\raise-.9ex\hbox to0pt{$\scriptstyle\sim\hss$}\hskip.8pt u{}}
\def\vtilde{\raise-.9ex\hbox to0pt{$\scriptstyle\sim\hss$}\hskip1pt \?v{}}
\def\nutilde{\raise-.9ex\hbox to0pt{$\scriptstyle\sim\hss$}\hskip1pt \?\nu{}}
\def\wtilde{\raise-.9ex\hbox to0pt{$\scriptstyle\sim\hss$} \?w{}}
\def\atilde{\raise-.9ex\hbox to0pt{$\scriptstyle\sim\hss$} a{}}
\def\btilde{\raise-.9ex\hbox to0pt{$\scriptstyle\sim\hss$} b{}}
\def\ftilde{\raise-1.4ex\hbox to0pt{$\scriptstyle\sim\hss$}\hskip1.3pt f{}}
\def\htilde{\raise-.9ex\hbox to0pt{$\scriptstyle\sim\hss$}\hskip.8pt h{}\hskip.2pt}
\def\ktilde{\raise-.9ex\hbox to0pt{$\scriptstyle\sim\hss$} k{}}
\def\ltilde{\hskip-1.2pt\raise-.9ex\hbox to0pt{$\scriptstyle\sim\hss$}\hskip3pt l{}}
\def\rtilde{\hskip-1pt\raise-1ex\hbox to0pt{$\scriptstyle\sim\hss$}\hskip2.2pt r{}\hskip-.6pt}
\def\stilde{\hskip-1pt\raise-1ex\hbox to0pt{$\scriptstyle\sim\hss$}\hskip1pt s{}}
\def\xitilde{\hskip-.5pt\raise-1.06ex\hbox to0pt{$\scriptstyle\sim\hss$}\hskip.5pt\xi{}}
\def\xtilde{\raise-.9ex\hbox to0pt{$\scriptstyle\sim\hss$}\hskip1.2pt x{}}
\def\ytilde{\raise-1.3ex\hbox to0pt{$\scriptstyle\sim\hss$}\hskip2pt y{}}
\def\ztilde{\hskip.4pt\raise-.9ex\hbox to0pt{$\scriptstyle\sim\hss$}\hskip1.7pt z{}}
\def\ztildecc{\hskip.4pt\raise-.9ex\hbox to0pt{$\scriptstyle\sim\hss$}\hskip1.7pt \overline z{}}

\def\C{\mathbb C}
\def\Chi{\text{\lower-2pt\hbox{$\chi$}}}

\def\N{\mathcal N}

\def\S{\mathcal S}
\def\P{\mathcal P}

\def\de#1/de#2{\frac{\partial{#1}}{\partial{#2}}}
\def\d#1/d#2{\frac{d#1}{d#2}}
\def\sd#1/de#2/de#3{\ifx#2 \frac{\plus02\partial^{\@\@2}#1}{\plus90\partial\@#3^{\@2}}%
\else\frac{\plus02\partial^{\@\@2}#1}{\partial\?#2\@\partial\?#3}\fi}
\def\oD#1/d#2{\textstyle{\text{\large$\d{#1}/d{#2}$}}}
\def\De#1/de#2{\textstyle{\text{\large$\de{#1}/de{#2}$}}}
\def\Sd#1/de#2/de#3{\textstyle{\text{\large$\sd{#1}/de{#2}/de{#3}$}}}

\def\TE{T\!\@S\!\@\?E\/(3)}

\def\Tr#1{{\plus60}^t\hskip-.8pt #1}
\def\Ttr#1{{\plus60}^t\hskip-1pt #1}
\def\TTr#1{{\plus60}^t\hskip-2pt #1}
\def\TTR#1{{\plus60}^t\hskip-3pt #1}

\def\@{\hskip.65pt}
\def\?{\hskip.3pt}
\def\plus#1#2{\vrule height#1pt width0pt depth#2pt}

\begin{document}

\title{Small oscillations of non-dissipative Lagrangian systems}

\author{Enrico Massa \and Stefano Vignolo}

\institute{E. Massa \at DIME, Sez.~Metodi e Modelli Matematici, Universit\`a di Ge\-no\-va. Via All'Opera Pia 15, 16145 Genova (Italy).  \\\email{massa@dima.unige.it}
    \and
          S. Vignolo \at DIME, Sez.~Metodi e Modelli Matematici, Universit\`a di Ge\-no\-va. Via All'Opera Pia 15, 16145 Genova (Italy). \\\email{vignolo@dime.unige.it}
}

\date{Received: date / Accepted: date}

\maketitle \abstract{The small oscillations of an arbitrary scleronomous systems subject to time-independent non dissipative forces are discussed. The linearized
equations of motion are solved by quadratures. As in the conservative case, the general integral is shown to consist of a superposition of harmonic oscillations. A
complexification of the resolving algorithm is presented.}


\date{}

\section{Introduction}
The concept of equilibrium stability, as well as the associated theory of \emph{small oscillations\/}, are standard topics in classical analytical mechanics.

Strangely enough, however, in the literature, with a few notable exceptions, the latter argument is dealt with under hypotheses definitely more restrictive than those
involved in the study of the former one: typically, while a sufficient condition for stability is established for arbitrary, time independent Lagrangians of the form
\begin{equation}\label{0.1}
L\@=\frac12\,a_{ij}\@\dot q^i\@\dot q^j+\@b_i\@\dot q^i +\@c:=L_2+L_1+L_0
\end{equation}
small oscillations are usually discussed under the simplifying assumption $\@L_1=0\@$ --- more specifically, assuming time--independent constraints and conservative
forces \cite{Whittaker,Levi,Finzi,Grioli,Fasano,Ruggeri}.

Possible extensions to non--conservative systems are also considered \cite{Gantmacher,Goldstein,Landau,Greenwood}, mainly in connection with the presence of dissipative
or gyroscopic effects.

In the present paper we propose an approach to the study of small oscillations for scleronomous systems
obeying the evolution equations
\begin{equation}\label{0.2}
\d/dt\,\de L/de{\dot q^k}\,-\,\de L/de{q^k}\,=\,Q_k
\end{equation}
with Lagrangian $\@L\/(q^h,\dot q^h)\@$ of the general form \eqref{0.1} and generalized forces $\@Q_k\/(q^h,\dot q^h)\@$ fulfilling the non--dissipa\-tivity condition
$\@Q_k\@\dot q^k=0\@$.

The resulting linearized equations of motion, viewed as first order differential equations in $\@\Re^{2\?n}$, are shown to be solvable by quadratures.

As expected, the general solution is a linear superposition of \emph{normal harmonics\/}, determined by the spectral structure of a symmetric, negative definite matrix,
expressing the square of the evolution operator.

An alternative characterization of the harmonics, based on a $n$-dimensional complex formalism is also worked out. The relation of the latter approach with the
$2\@n$--dimensional, real one is discussed.

\section{Small oscillations}
Under the assumptions stated in the Introduction, every strict local maximum $q^*=(q^{*\@1}\And q^{*\@n})$ of the function $\@L_0\@$ is readily seen to represent a
stable equilibrium configuration for the given system \cite{Greenwood}.

In fact, in view of the condition $\@Q_r\@\dot q^r=0\@$ and of the consequent relation
\begin{equation}\label{2.1}
0=\de/de{\dot q^k}\,(Q_r\@\dot q^r)\@=\@Q_k+\@\de Q_r/de{\dot q^k}\,\dot q^r
\end{equation}
the generalized forces do not play any role in the determination of the equilibrium configurations, nor in the applicability of Dirichlet's stability theorem.

The second--order approximation of the La\-grangian in a neighborhood of the kinetic state $\@(q^*,0)\@$ reads
\begin{equation*}
\tilde L\,=\@\tfrac12\,A_{kr}\,\dot\eta^k\dot\eta^r+ \Big[b_r\/(q^*)\@+\@\Big(\de b_r/de{q^k}\Big)_{\!q^*}\@\eta^k \Big]\@\dot\eta^r
-\@\tfrac12\,C_{kr}\,\eta^k\eta^r
\end{equation*}
with $A_{kr}=a_{kr}\/(q^*)$, $C_{kr}=-\Big(\Sd L_0/de{q^k}\!/de{q^r}\Big)_{\!q^*}$, $\eta^k=q^k-q^{*\?k}\!$.

From the latter we get the expressions
\begin{multline*}
\d/dt\,\de\tilde L/de{\dot \eta^k}\,-\,\de\tilde L/de{\eta^k}=                                                                            \\
=A_{kr}\@\ddot\eta^r\@+\left[\Big(\de b_k/de{q^r}\Big)_{\!q^*}-\Big(\de b_r/de{q^k}\Big)_{\!q^*}\right]\dot\eta^r+ C_{kr}\@\eta^r
\end{multline*}

In a similar way, on account of eq.~\eqref{2.1}, the first order approximation of the generalized forces in a neighborhood of $\@(q^*,0)\@$ takes the form
\begin{equation*}
\tilde Q_k= \biggl(\de Q_k/de{\dot q^r}\biggr)_{\!(q^*\!,0)}\dot\eta^r=-\biggl(\de Q_r/de{\dot q^k}\biggr)_{\!(q^*\!,0)}\dot\eta^r
\end{equation*}

Collecting all results, and setting
\begin{equation*}
B_{kr}:=\@\Big(\de b_k/de{q^r}\Big)_{\!q^*}-\Big(\de b_r/de{q^k}\Big)_{\!q^*}-\Big(\de Q_k/de{\dot q^r}\Big)_{\!(q^*\!,0)}
\end{equation*}
the linearized equations of motion read
\begin{equation*}
A_{kr}\@\ddot\eta^r\@+\@B_{kr}\@\dot\eta^r\@+\@C_{kr}\@\eta^r=\,0
\end{equation*}
or, synthetically
\begin{equation}\label{2.2}
A\@\ddot\etatilde\,+\,B\@\dot\etatilde\,+\,C\@\etatilde\,=\,0
\end{equation}
where $A$ and $C$ are symmetric, positive definite matrices, $B$ is an antisymmetric one, and $\etatilde $\vspace{-3pt} is the column vector
$\@\begin{pmatrix}\eta^1\\[-2.5pt] :\; \\[-3pt] \eta^n\end{pmatrix}\@$.\vspace{2pt}

Viewed as a system of first order ODE in the velocity space, eq.~\eqref{2.2} may be written in the normal form
\begin{equation*}
\left\{\,
\begin{aligned}
 & \d\etatilde/dt \,=\,\dot\etatilde \\[2pt]
 &\d\dot\etatilde/dt \,=\,-\@A^{-1}\@C\@\etatilde\,-\@A^{-1}\@B\@\dot\etatilde
\end{aligned}
\right.
\end{equation*}
or, in matrix notation
\begin{equation*}\tag{\ref{2.2}'}
\d/dt
\begin{pmatrix}
  \etatilde \\
  \@\dot\etatilde\@
\end{pmatrix}=
\begin{pmatrix}
  0&\;\,I \\[2pt]
  -\@A^{-1}\@C& \;\,-\@A^{-1}\@B
\end{pmatrix}
\begin{pmatrix}
  \etatilde \\
  \@\dot\etatilde\@
\end{pmatrix}:=M\begin{pmatrix}
  \etatilde \\
  \@\dot\etatilde\@
\end{pmatrix}
\end{equation*}

The non-singular endomorphism $M:\Re^{2\?n}\to\Re^{2\?n}$ described by the matrix
\begin{equation}\label{2.3}
M\,=\,\begin{pmatrix}
  0&\;\;I \\[2pt]
  -\@A^{-1}\@C& \;\;-\@A^{-1}\@B
\end{pmatrix}
\end{equation}
will be called the \emph{linearized evolution operator\/}.

In addition to $M\@$, another important operator is the bilinear functional $\@\Re^{2\?n}\times\Re^{2\?n}\to\Re\@$ sending each pair of vectors $\@\utilde,\vtilde\@$
into the scalar $\@\Tr\utilde\@ K\vtilde\@$, $\@K\@$ denoting the non-singular, antisymmetric matrix
\begin{equation}\label{2.4}
K\,:=\,\begin{pmatrix}
  \;B&\;A \\
  -\@A&\;0
\end{pmatrix}
\end{equation}

For clarity, a few comments are in order:
\begin{itemize}
\item
linear endomorphisms and bilinear functionals are of course \emph{different\/} mathematical objects, with different composition rules: for example, in the present
context, a formally legitimate expression like $\Tr\utilde\?M\vtilde\@$ or $\@K^2\@$ has no invariant geometrical meaning, while $\@\Tr\utilde\, K M\@\vtilde\@$ or
$\@M^{\?2}\@$ are perfectly significant ones;\vspace{1pt}
\item
every symmetric positive definite bilinear functional may be used to define a \emph{scalar product\/} over $\@\Re^{2\?n}\@$;\vspace{1pt}
\item
a linear endomorphisms $\@\psi:\Re^{2\?n}\to\Re^{2\?n}\@$ is symmetric (antisymmetric) with respect to a given scalar product $\@(\;,\;)\@$ if and only if the
expression $\@(\utilde\@,\@\psi\/\vtilde)\@$ is symmetric (respectively antisymmetric) in the arguments $\@\utilde,\vtilde\@$. In particular, if $\@\psi\@$ is
symmetric, the space $\@\Re^{2\?n}\@$ admits an orthonormal basis formed by eigenvectors of $\@\psi\@$.

\end{itemize}

With this in mind, let us now observe the following basic facts:
\begin{itemize}
\item
the bilinear functional associated with the matrix $\@K M^{-1}\@$ is symmetric and positive definite. The conclusion follows at once from the identities
\begin{equation*}
\Tr K M=\begin{pmatrix}
  -\@B&-\@A \\
  \@A&0
\end{pmatrix}\!
\begin{pmatrix}
  0&\;I \\
  -\@A^{-1}\@C&\;-\@A^{-1}\@B
\end{pmatrix}=
\begin{pmatrix}
 \@C&\@0 \\
 \@ 0&\@A
\end{pmatrix}
\end{equation*}
\begin{equation*}
K M^{-1}=\,K\,\big[\@\big(\Tr K M\big){}^{-1}\@\big]\,\Tr K\,=\,K
\begin{pmatrix}
 \@C^{-1}&\@0 \\
 \@ 0&\@A^{-1}
\end{pmatrix} \Tr K
\end{equation*}
showing that $\,K\cdot M^{-1}\@$ is congruent to a symmetric, positive definite matrix;
\item
the operator $\@M\@$ is \emph{antisymmetric\/} with respect to the scalar product defined by the prescription
\begin{equation}\label{2.5}
(\utilde\,,\?\vtilde):=\@\Tr\utilde\,K M^{-1}\vtilde
\end{equation}
Indeed, from eq.~\eqref{2.5} and the antisymmetry of $\@K\@$, it directly follows $\@(\utilde\,,M\vtilde)=\@\Tr\utilde\,K\@\vtilde=-\@(\vtilde\,,M\?\utilde)\@$;
\item
the operator $M^{\?2}$ is symmetric and negative definite with respect to the scalar product \eqref{2.5}: the antisymmetry and non--singularity of $M$ entail in fact
the relations
\begin{align*}
&(\@\utilde\,,M^{\?2}\@\vtilde\,)=-\,(M\utilde\,,\@M\vtilde\,)=(M^{\?2}\utilde\,,\vtilde\@)                                   \\
&(\@\utilde\,,\@M^{\?2}\utilde\,)=-\,(M\utilde\,,\@M\utilde\,)<0 \quad\forall\,\utilde\in\Re^{2\?n},\,\utilde\ne 0
\end{align*}
\end{itemize}

After these preliminaries, we now state
\begin{theorem}\label{Teo2.1}
The vector space $\@\Re^{2\?n}$ admits at least one basis $\@\utilde_{\?k},\vtilde_{\?k}\,,\;k=1\And n\@$, orthonormal with respect to the scalar product \eqref{2.5} and
satisfying the relations
\begin{equation}\label{2.6}
M\?\utilde_{\@k}\@=\@\w_k\@\vtilde_{\?k}\,,\quad M\?\vtilde_{\?k}\@=\@-\,\w_k\@\utilde_{\@k}
\end{equation}
with $\@\w_1\And\w_n\@$ positive, not necessarily distinct real numbers.
\end{theorem}
\begin{Proof}
Let $\@-\@\l_\a^{\;2}\@$, $\@\l_\a\in\Re_+\@$, $\@\a=1\And r\@$ denote the distinct eigenvalues of $\@M^{\?2}$, and $\@\S_\a\subset \Re^{2\?n}\@$ the corresponding
eigenspaces. To each $\@\utilde\in\S_\a\@$ we associate a ``partner vector'' $\@\vtilde:=\l_\a^{-1}\@M\utilde\@$.
The resulting pair $\@\utilde\@,\vtilde\@$ satisfies then the relations
\begin{align*}
& M\utilde=\l_\a\,\vtilde\,;\; M\vtilde=\l_\a^{-1}\@M^{\?2}\utilde\@=-\@\l_\a\,\utilde\,;                       \\
& M^{\?2}\vtilde\@=-\@\l_\a\@M\utilde=-\@\l_\a^2\@\vtilde\,;                                                    \\
&(\@\utilde,\vtilde\@)\@=\@\l_\a^{-1}\@(\@\utilde,\@M\?\utilde\@)\@=0\,;                                        \\
&(\@\vtilde,\vtilde\@)\@=\@\l_\a^{-2}\@(M\?\utilde,\@M\?\utilde\@)\@=\@
-\,\l_\a^{-2}\@(\@\utilde,\@M^{\?2}\utilde\@)\@=\@(\@\utilde,\utilde\@)\,.
\end{align*}

These show that, like $\@\utilde\@$, the vector $\vtilde$ belongs to the eigenspace $\S_\a$, and that $\utilde$, $\vtilde$ are mutually orthogonal, have the same norm,
and fulfil an equation of the form \eqref{2.6}, with $\@\w_k=\l_\a$
\footnote%
{In particular the dimension of each $\S_\a$ is necessarily an even number, henceforth denoted by $2\@n_\a\@$.}.

Moreover, setting $\@V:=\Span\/(\utilde\@,\vtilde)\@$, the relations
\begin{align*}
& (\@M\wtilde,\utilde\@)=-(\@\wtilde,M\utilde\@)=-\@\l_\a\@(\@\wtilde,\vtilde\@)                                 \\
& (\@M\wtilde,\vtilde\@)=-(\@\wtilde,M\vtilde\@)=\@\l_\a\@(\@\wtilde,\utilde\@)
\end{align*}
show that the operator $\@M\@$ maps the subspace $\@V^\perp\@$ onto itself, thereby inducing a non--singular antisymmetric endomorphism $\@M_{|\?V^\perp}:V^\perp\to
V^\perp\@$, whose square is of course identical to the restriction $\@M^{\?2}_{\;|\?V^{\!\perp}}\@$

The rest of the proof proceeds by induction: for $n~=~1$ (namely in $\Re^2$), choosing $\utilde$ of unit norm and setting $\@\utilde_{\@1}=\utilde$,
$\,\vtilde_{\@1}=\vtilde$, $\,\w_1=\l\@$ establishes the thesis.

In a similar way, for $\@n>1\@$, we arbitrarily select an eigenspace $\@\S_\a\@$ of $\@M^2\@$, a unit vector $\@\utilde_{\@n}\in\S_\a\@$, and denote by
$\@\vtilde_{\@n}=\l_\a^{-1}M\utilde_{\@n}\@$ the associated partner vector.

The thesis then follows setting $\@V_n=\Span\/(\utilde_{\@n}\@,\vtilde_{\@n}\!)\@$, $\@\w_n=\l_\a\@$, and applying the inductive hypothesis to the
$\@(2\?n-2)$--dimensional subspace $\@V_n^\perp\@$.
\end{Proof}

Introducing the notation
\begin{equation}\label{2.7}
\utilde_k=\begin{pmatrix}\htilde_k\\[1pt]\ltilde_k\end{pmatrix},\;\vtilde_k=\begin{pmatrix}\rtilde_k\\[1pt]\stilde_k\end{pmatrix}\quad\;\big(\htilde_k\?,\@\ltilde_k\?,
\@\rtilde_k\?,\@\stilde_k\in\Re^n\big)
\end{equation}
eqs.~\eqref{2.3}, \eqref{2.6} imply the equalities
\begin{align*}
& \begin{pmatrix}
0&\quad I                       \\[2pt]
-\@A^{-1}\@C& \;\;-\@A^{-1}\@B
\end{pmatrix}
\begin{pmatrix}
\htilde_{\@k}                   \\[2pt]
\ltilde_{\@k}
\end{pmatrix}
=\;\;\w_k
\begin{pmatrix}\rtilde_{\@k}    \\[2pt]
\stilde_{\@k}
\end{pmatrix}
                                \\[4pt]
& \begin{pmatrix}
0&\quad I                       \\[2pt]
-\@A^{-1}\@C& \;\;-\@A^{-1}\@B
\end{pmatrix}
\begin{pmatrix}
\rtilde_{\@k}                   \\[2pt]
\stilde_{\@k}
\end{pmatrix}
=-\,\w_k\!
\begin{pmatrix}\htilde_{\@k}    \\[2pt]
\ltilde_{\@k}
\end{pmatrix}
\end{align*}
summarized into the pair of real relations
\begin{equation}\label{2.8}
\ltilde_{\@k}\@=\@\w_k\,\rtilde_{\@k}\@,\qquad \stilde_{\@k}\@=\@-\,\w_k\,\htilde_{\@k}
\end{equation}
completed by the complex one
\begin{equation}\label{2.9}
(C-\@i\@\w_k\@B-\@\w_k^{\;2}\@A)(\htilde_{\@k}+i\@\rtilde_{\@k})\,=\,0
\end{equation}

Referring $\@\Re^{2\?n}\@$ to the basis $\@\utilde_k,\@\vtilde_k\@$ indicated in Theorem \ref{Teo2.1}, and putting
\begin{equation*}
\begin{pmatrix}
  \etatilde \\
  \,\dot\etatilde\,
\end{pmatrix}\@=\,\Chi_k\/(t)\,\utilde_{\@k}\,+\,\psi_k\/(t)\@\vtilde_k
\end{equation*}
we rewrite the evolution equations (\ref{2.2}') in the form
\begin{align*}
\dot\Chi_k\/(t)\,\utilde_{\@k}\,+\,\dot\psi_k\/(t)\,\vtilde_k\,=\,&M\?\big[\@\Chi_k\/(t)\,\utilde_{\@k}\,+\,\psi_k\/(t)\,\vtilde_k\@\big]\@=         \\[2pt]
&=\,\w_k\?\big[\@\Chi_k\/(t)\,\vtilde_k\,-\,\psi_k\/(t)\,\utilde_{\@k}\@\big]
\end{align*}
mathematically equivalent to the system
\begin{equation*}
\left\{\,
\begin{aligned}
 & \dot\Chi_k\,=\,-\,\w_k\,\psi_k \\
 & \dot\psi_k\,=\;\;\;\w_k\,\Chi_k
\end{aligned}\right.\hskip2cm\text{(not summed over $\@k\@$)}
\end{equation*}

The latter admits the general integral
\begin{equation*}
\Chi_k\,=\,a_k\@\cos\?(\w_k\@t+\vphi_k)\,,\qquad \psi_k\,=\,a_k\@\sin\?(\w_k\@t+\vphi_k)
\end{equation*}
with $\@a_k\@,\,\vphi_k\@$ arbitrary constants.

The solution of eqs.~(\ref{2.2}') is therefore
\begin{equation*}
\begin{pmatrix}
  \etatilde \\
  \,\dot\etatilde\,
\end{pmatrix}\@=\@\sum_{k=1}^n\,a_k\@\big[\@\cos\/(\w_k\@t+\vphi_k)\,\utilde_{\@k}\,+\,\sin\/(\w_k\@t+\vphi_k)\,\vtilde_k\,\big]
\end{equation*}

Recalling eqs.~\eqref{2.7}, we conclude that the general integral of the linearized equations of motion \eqref{2.2} reads
\begin{equation}\label{2.10}
\etatilde\@=\@\sum_{k=1}^n\,a_k\@\big[\cos\/(\w_k\@t+\vphi_k)\,\@\htilde_{\@k}\,+\,\sin\/(\w_k\@t+\vphi_k)\,\@\rtilde_k\@\big]
\end{equation}
while the expression for $\@\dot\etatilde\@$ coincides with the one obtained from eq.~\eqref{2.10}, taking the identifications \eqref{2.8} into account.

As expected, the motion of the system consists of a linear superposition $\@\etatilde=\sum_k\?a_k\,\nutilde_k\@$ of harmonic oscillations, henceforth called the
\emph{normal harmonics\/}.

The determination of the general integral
\eqref{2.10} can be simplified by replacing the $2\@n$-dimensional real formalism with a $n$-dimensional \emph{complex\/} one.

To this end, resuming the notations adopted in the proof of Theorem \ref{Teo2.1}, we indicate by $\@\S_\a\@$, $\@\a=1\And r\@$ the eigenspaces of the operator
$\@M^{\?2}$, by $\@-\@\l_\a^{\;2}\@$ ($\@\l_\a>0\@$) the corresponding eigenvalues, and by $2\@n_\a=\dim\/(\S_\a)$ the respective multiplicities.

Also, for each $\@\a$, we denote by $\@\N_\a\@$ the kernel of the endomorphism $\@(C-i\@\l_\a\@B-\l_\a^2\@A):\C^n\to\C^n$.

We have then the following
\begin{theorem}\label{Teo2.2}
The space $\@\C^n\@$ splits into the direct sum
\begin{equation}\label{2.11}
\C^n=\@\oplus_{\a=1}^r\,\N_\a
\end{equation}
with $\@dim\/(\N_\a)\@=\@n_\a\@$
\end{theorem}
\begin{Proof}
To start with, let us establish the intersection property $\@\N_\a\cap\N_\b\@=\@\{0\}\;\,\forall\,\a\ne\b\@$.

To this end we observe that, for any $\@\ztilde\in\N_\a\cap\N_\b\@$, the simultaneous validity of the conditions
\begin{subequations}\label{2.12}
\begin{align}
 & (C-\@i\,\l_\a\@\@B-\,\l_\a^{\,2}\@A)\@\ztilde\,=\,0                \\[2pt]
 & (C-\@i\,\l_\b\@\@B-\,\l_\b^{\,2}\@A)\@\ztilde\,=\,0
\end{align}
\end{subequations}
implies the relation
\begin{multline*}
\big[\@i\@(\l_\a-\l_\b)\@B\@+\@(\l_\a^{\;2}-\l_\b^{\;2})\@A\@\big]\@\ztilde\@=                 \\[2pt]
=\@(\l_\a-\l_\b)\@\big[\@i\@B\@+\@(\l_\a+\l_\b)\@A\@\big]\@\ztilde\@=\@0      \hskip1cm
\end{multline*}
whence, dividing by $\l_\a-\l_\b$ and substituting into equation (\ref{2.12}\@a)
\begin{equation*}
\big[\@C\@+\@\l_\a\@(\l_\a+\l_\b)\@A\@-\@\l_\a^{\;2}\@A\@\big]\@\ztilde\@=(C+\l_\a\@\l_\b\@A)\@\ztilde\@=\@0
\end{equation*}

At the same time, the positiveness of $\l_\a\@,\l_\b\@$, together with the positive definiteness of $A$ and $C\@$, ensure the non--singularity of the matrix
$\@C+\l_\a\@\l_\b\@A\@$. Therefore, $\@\ztilde\in\N_\a\cap\N_\b\,\Longleftrightarrow\,\ztilde=0\@$.\vspace{1pt}

Let us now evaluate the \emph{dimension\/} of each $\@\N_\a\@$. To this end, referred $\@\Re^{2n}\@$ to an orthonormal basis satisfying eqs.~\eqref{2.6}, we split the
vectors $\@\utilde_k, \vtilde_k\@$ into $r$ distinct subfamilies $\@\big\{\utilde^{(\a)}_{\@i_\a}\!,\vtilde^{(\a)}_{\@i_\a}\!,\;i_\a=1\And n_\a\@\big\}$\vspace{1pt},
each one spanning a corresponding eigenspace $\@\S_\a\@$.

Restoring the notation \eqref{2.7}, to each pair of ``partner vectors'' $\@\utilde^{(\a)}_{\@i_\a}\!,\vtilde^{(\a)}_{\@i_\a}\@$ we associate the complex vector
$\@\ztilde^{(\a)}_{\@i_\a}= \htilde^{(\a)}_{\@i_\a}\@+i\@\rtilde^{(\a)}_{\@i_\a}\in\C^n$.

In view of eq.~\eqref{2.9}, $\@\ztilde^{(\a)}_{\@i_\a}\@$ belongs to the kernel $\@\N_\a\@$. Moreover, the vectors $\@\big\{\ztilde^{(\a)}_{\@i_\a},\;i_\a=1\And
n_\a\big\}\@$\vspace{1pt} are linearly independent: due to the stated definitions, every equation of the form
$\@\sum_{i_\a}\@(a_{\@i_\a}+i\,b_{\@i_\a})\@\ztilde^{(\a)}_{\@i_\a}\!=0$\vspace{2pt} splits in fact into the pair of conditions
\begin{equation*}
\sum_{i_\a}\@\big(a_{\@i_\a}\,\htilde^{(\a)}_{\@i_\a}+b_{\@i_\a}\,\rtilde^{(\a)}_{\@i_\a}\big)=
\sum_{i_\a}\@\big(a_{\@i_\a}\,\rtilde^{(\a)}_{\@i_\a}-b_{\@i_\a}\,\htilde^{(\a)}_{\@i_\a}\big)=0
\end{equation*}
which, summarized into the single expression
\begin{equation*}
\sum_{i_\a}\@\big(a_{\@i_\a}\,\utilde^{(\a)}_{\@i_\a}+b_{\@i_\a}\,\vtilde^{(\a)}_{\@i_\a}\big)=0
\end{equation*}
ensure the vanishing of all coefficients $\@a_{\@i_\a},b_{\@i_\a}\@$.

The dimension of $\@\N_\a\@$ is therefore \emph{not less\/} than $\@n_\a\@$. Being $\@\sum_\a \@n_\a = n\@$, this fact, together with the intersection property,
establishes the thesis.
\end{Proof}

According to Theorem \ref{Teo2.2}, the space $\@\C^n\@$ admits at least one basis $\@\big\{\ztilde_1\And\ztilde_n\big\}\@$ whose elements satisfy equations of the form
\begin{equation}\label{2.13}
(C-\@i\,\w_k\@\@B-\,\w_k^{\;2}\@A)\ztilde_k\,=\,0
\end{equation}
$\@\w_k\@$ being $n$ (not necessarily distinct) \emph{positive\/} roots of the equation $\@\det(C-\@i\,\w\?B-\,\w^2\/A)=0\@$.

Eq.~\eqref{2.13} implies the differential relation
\begin{equation*}
\biggl(A\,\d{\@\plus50}^2/d{t^2}\,+\,B\,\d/dt\,+\,C\biggr)\big(e^{-i\@\w_k\@t}\@\ztilde_k\big)\,=\,0
\end{equation*}
and the corresponding complex conjugate
\begin{equation*}
\biggl(A\,\d{\@\plus50}^2/d{t^2}\,+\,B\,\d/dt\,+\,C\biggr)\big(e^{i\@\w_k\@t}\@\ztildecc_k\big)\,=\,0
\end{equation*}

In the space $\@\C^n\@$, the second order differential equation $\@A\,\ddot{\!\ztilde}\@+\@B\@\dot{\!\ztilde}\@+\@C\ztilde=0\@$ for the unknown
$\@\ztilde=\ztilde\/(t)\@$ admits therefore the general integral
\begin{equation}\label{2.14}
\ztilde\@=\@\sum_{k=1}^n\big(\g_k\,e^{-\@i\@\w_k\@t}\ztilde_k\@+\@\delta_k\,e^{i\@\w_k\@t}\@\ztildecc_k\big)
\end{equation}
$\@\g_k\?,\?\delta_k\,$ being arbitrary complex constants.

The right-hand-side of \eqref{2.14} is \emph{real\/} if and only if $\@\delta_k=\overline{\plus50\g}_k\@$. Setting $\@\g_k\@=\@a_k\,e^{-\?i\@\vphi_k}\@$\vspace{1pt} and
restoring the notation $\@\ztilde_k=\htilde_k+i\@\rtilde_k\,$\vspace{1pt}, we obtain in this way the general integral of equation \eqref{2.2} in the form
\begin{equation*}
\etatilde\/(t)\@=\@\Real\,\sum_{k=1}^n\left[\,a_k\,e^{-\?i\?(\w_k\@t+\vphi_k)}\@\big(\htilde_k+i\,\rtilde_k\big)\,\right]
\end{equation*}
clearly identical to eq.~\eqref{2.10}.

\smallskip
Although not directly relevant to the implementation of the algorithm, it may be noticed that, unlike what happens in $\@\Re^{2n}$, the characterization of the bases
$\big\{\ztilde_1\And\ztilde_n\big\}$ of $\C^n\@$ fulfilling the requirements \eqref{2.13} does not involve any concept of orthonormality.

This lack of symmetry between the real and the complex formalism may be disposed of by endowing each subspace $\@\N_\a\subset\C^n\@$ with a sesquilinear scalar product,
based on the prescription
\begin{equation}\label{2.15}
(\ztilde,\@\wtilde)\,=\,\TTr\ztildecc\,\biggl(A\@+\,\frac{i\@B}{2\@\l_\a^{\,2}}\biggr)\,\wtilde
\end{equation}

The definition of $\@\N_\a\@$ entails in fact the relation
\begin{multline*}
\TTr\ztildecc\@\biggl(A\@+\,\frac{i\@B}{2\@\l_\a}\biggr)\@\wtilde\@=\@
\frac1{2\@\l_\a^{\,2}}\;\,\TTr\ztildecc\,\@\big(2\@\l_\a^{\,2}\,A\@+\@\l_\a\@i\@B\big)\,\wtilde\@=                                           \\[2pt]
=\@\frac1{2\@\l_\a^{\,2}}\;\,{\plus70}^t\!\@\ztildecc\,\@\big(\l_\a^{\,2}\,A\@+\@C\big)\,\wtilde
\end{multline*}
ensuring the positiveness of $\@(\ztilde,\ztilde)\@$ $\,\forall\,\ztilde\in\N_\a\@$.

The various prescriptions \eqref{2.15} can be glued into a single scalar product in $\@\C^n$, by adding the requirement of \emph{mutual orthogonality\/} of the kernels
$\@\N_\a\@$.

Denoting by $\@\P_\a:\C^n\to\N_\a\@$ the family of projections associated with the direct sum decomposition \eqref{2.11}, this leads to the expression
\begin{equation*}
(\ztilde,\@\wtilde)\@=\@\sum_{\a=1}^r\,
{\plus80}^t\!\@\big(\,\overline{\plus70\P_\a\@\ztilde}\,\big)\@\biggl(A\@+\,\frac{i\@B}{2\@\l_\a^{\,2}}\biggr)\@\P_\a\@\wtilde
\quad\forall\,\ztilde,\wtilde\in\C^n
\end{equation*}

We have then the following
\begin{theorem}\label{Teo2.3}
The orthonormal bases of $\Re^{2n}$ fulfilling the requirement \eqref{2.6} are in 1--1 correspondence with the orthonormal bases of $\@\C^n\@$ fulfilling the requirement
\eqref{2.13}.
\end{theorem}
\begin{Proof}
Due to the orthogonal character of both direct sum decompositions $R^{2\?n}=\oplus_\a\@\S_\a\@$ and $\@\C^n=\oplus_\a\@\N_\a\@$, it is sufficient to discuss the relation
between bases in $\@\S_\a\@$ and bases in $\@\N_\a\@$.

To this end, recalling eqs.~\eqref{2.7}, \eqref{2.8} and adapting the notation, we represent  each pair of partner vectors of the basis $\big\{\@\utilde^{(\a)}_{\@i_\a},
\vtilde^{(\a)}_{\@i_\a}\@,\;\a=1\And n_\a\big\}\@$ in the form
\begin{equation}\label{2.16}
\utilde^{(\a)}_{\@i_\a}=\begin{pmatrix}\@\xtilde^{(\a)}_{\@i_\a}\\[2pt]\,\l_\a\@\ytilde^{(\a)}_{\@i_\a}\end{pmatrix},\quad
\vtilde^{(\a)}_{\@i_\a}=\begin{pmatrix}\@\ytilde^{(\a)}_{\@i_\a}\\[2pt]-\@\l_\a\@\xtilde^{(\a)}_{\@i_\a}\end{pmatrix}
\end{equation}

From the proof of Theorem \ref{Teo2.2} we know that the complex vectors $\ztilde^{(\a)}_{\@i_\a}=\xtilde^{(\a)}_{\@i_\a}+i\@\ytilde^{(\a)}_{\@i_\a}\@$ span $\@\N_\a\@$:
all we have to do is therefore to check that the correspondence $\utilde^{(\a)}_{\@i_\a}\!, \vtilde^{(\a)}_{\@i_\a}\to\ztilde^{(\a)}_{\@i_\a}$ preserves the
orthonormality relations. And, in fact:
\\[2pt]
-- the definition of $\@\N_\a\@$ as the kernel of the operator $\@C-i\@\l_\a\@B-\l_\a^2\@A\@$ implies the identities
\begin{equation*}
\TTR\ztilde^{(\a)}_{\@i_\a}\@\big(C-i\@\l_\a\@B-\l_\a^{\,2}\@A\big)\@\ztilde^{(\a)}_{\@j_\a}\,=\@0\quad \forall\,i_\a,\@j_\a=1\And n_\a
\end{equation*}

Due to the symmetry properties of the matrices $\@A,B,C$, the latter split into the pair of expressions
\begin{equation*}
\TTR\ztilde^{(\a)}_{\@i_\a}\@\big(C-\l_\a^{\,2}\@A\big)\@\ztilde^{(\a)}_{\@j_\a}\,=\@0\,,\qquad
\TTR\ztilde^{(\a)}_{\@i_\a}\@B\@\ztilde^{(\a)}_{\@j_\a}\,=\@0
\end{equation*}

From the second one, setting $\ztilde^{(\a)}_{\@i_\a}=\xtilde^{(\a)}_{\@i_\a}+i\@\ytilde^{(\a)}_{\@i_\a}$, we get the equalities
\begin{align*}
&\Ttr\xtilde^{(\a)}_{\@i_\a}\@B\@\xtilde^{(\a)}_{\@j_\a}\,=\@\TTr\ytilde^{(\a)}_{\@i_\a}\@B\@\ytilde^{(\a)}_{\@j_\a}           \\
&\Ttr\xtilde^{(\a)}_{\@i_\a}\@B\@\ytilde^{(\a)}_{\@j_\a}\@+\@\TTr\ytilde^{(\a)}_{\@i_\a}\@B\@\xtilde^{(\a)}_{\@j_\a}\,=\@0
\end{align*}

In view of these, taking eq.~\eqref{2.15} into account, it is readily seen that the conditions $\@(\ztilde_{\@i_\a},\ztilde_{\@j_\a})=\delta_{{\@i_\a\?j_\a}}\@$ are
expressed by the equations
\begin{subequations}\label{2.17}
\begin{align}
& \Ttr\xtilde_{\?i_\a}A\@\xtilde_{\?j_\a} +\@\TTr\ytilde_{\?i_\a}A\@\ytilde_{\?j_\a}-\@
\frac1{\l_\a}\;\Ttr\xtilde_{\?i_\a}B\@\ytilde_{\?j_\a}\@=\delta_{{\@i_\a\?j_\a}}                                    \\[2pt]
& \Ttr\xtilde_{\@i_\a}A\@\ytilde_{\@j_\a}\@-\@\TTr\ytilde_{\@i_\a}A\@\xtilde_{\@j_\a}\@+\@
\frac1{\l_\a}\;\Ttr\xtilde_{\@i_\a}B\@\xtilde_{\@j_\a}\@=\@0
\end{align}
\end{subequations}

\noindent
 -- taking eqs.~\eqref{2.6} and the properties of the operator $\@M\@$ into account, the scalar products between the basis vectors in $\@\S_\a\@$ read
\begin{align*}
& \big(\utilde^{(\a)}_{\@i_\a},\utilde^{(\a)}_{\@j_\a}\big)=-\frac1{\l_\a}\,\big(\utilde^{(\a)}_{\@i_\a},M\vtilde^{(\a)}_{\@j_\a}\big)
=-\frac1{\l_\a}\;\Ttr\utilde^{(\a)}_{\@i_\a}\@K\@\vtilde^{(\a)}_{\@j_\a}                                                                \\[2pt]
& \big(\utilde^{(\a)}_{\@i_\a},\vtilde^{(\a)}_{\@j_\a}\big)\,=\;\frac1{\l_\a}\,\big(\utilde^{(\a)}_{\@i_\a},M\utilde^{(\a)}_{\@j_\a}\big)
\,=\,\frac1{\l_\a}\;\Tr\utilde^{(\a)}_{\@i_\a}\@K\@\utilde^{(\a)}_{\@j_\a}                                                                 \\[2pt]
& \big(\vtilde^{(\a)}_{\@i_\a},\vtilde^{(\a)}_{\@j_\a}\big)=\@\frac1{\l_\a^2}\,\big(M\utilde^{(\a)}_{\@i_\a},M\utilde^{(\a)}_{\@j_\a}\big)
=\@\big(\utilde^{(\a)}_{\@i_\a},\@\utilde^{(\a)}_{\@j_\a}\big)
\end{align*}

In view of eqs.~\eqref{2.5}, \eqref{2.16}, the orthonormality relations have therefore the form
\begin{align*}
 -&\frac1{\l_\a}\@\big(\@\TTr\xtilde_{\@i_\a},\l_\a\,\TTr\ytilde_{j_\a}\big)
\begin{pmatrix}
  \;B&\;A \\[2pt]
  -\@A&\;0
\end{pmatrix}
\begin{pmatrix}
\ytilde_{\@j_\a}\plus06\\[2pt]
-\l_\a\@\xtilde_{\@j_\a}
\end{pmatrix}
= \,\delta_{i_\a j_\a}                                                                                                                      \\[3pt]
& \frac1{\l_\a}\@\big(\@^t\!\xtilde_{\@i_\a},\l_\a\@^t\!\ytilde_{\@j_\a}\big)
\begin{pmatrix}
  \;B&\;A \\[2pt]
  -\@A&\;0
\end{pmatrix}
\begin{pmatrix}
\xtilde_{\@j_\a}\plus06\\[2pt]
\l_\a\@\ytilde_{j_\a}
\end{pmatrix}
= \,0
\end{align*}
identical to eqs.~(\ref{2.17}\@a,b).
\end{Proof}

\begin{remark}\label{Rem2.1}
A deeper insight into the geometrical content of Theorem \ref{Teo2.3} is gained by considering the totality of $\@2$--dimensional subspaces $\@\Sigma\subset\S_\a\@$
\emph{invariant\/} under the action of the operator $\@M\@$.

Each such $\@2$--space is completely determined by the knowledge of any of its non-zero elements $\@\utilde\@$, through the identification
$\@\Sigma=\Span\/(\utilde,M\utilde)\@$

Preserving the notation of Theorem \ref{Teo2.3}, let us now denote by $\psi_\a:\S_\a\to\N_\a$\vspace{1pt} the correspondence sending each vector
$\utilde=\bigg(\begin{matrix}\xtilde\\[2pt]\l_\a\?\ytilde\end{matrix}\bigg)$ into the image $\psi_\a\/(\utilde)=\xtilde+i\ytilde$.\vspace{-3pt}

Due to the relation $\@M\utilde=\l_\a\bigg(\begin{matrix}\ytilde\\[2pt]\,-\@\l_\a\@\xtilde\end{matrix}\bigg)$,\vspace{2pt} we have then
$\@\psi_\a\/(M\utilde)=\l_\a\@(\ytilde-i\xtilde)=-i\@\l_\a\@\psi_\a\/(\utilde)\@$\vspace{1pt}: the image $\@\psi_\a\/(\Sigma)\@$ of an invariant $2$--plane in
$\S_\a$\vspace{1pt} is therefore a \emph{direction\/} in $\@\N_\a\@$.

In connection with this mental picture, Theorems \ref{Teo2.1}, \ref{Teo2.3} point out the following features:
\begin{itemize}
\item the map $\@\psi_\a\@$ sets up a 1-1 correspondence between mutually orthogonal invariant $2$--planes and mutually orthogonal directions;
\item each invariant $2$--plane $\@\Sigma\subset \S_\a$ admits a $1$--parame\-ter family of orthonormal bases $\@\utilde,\vtilde\@$, defined up to an arbitrary rotation
\begin{equation*}
\left\{
\begin{aligned}
& \utilde'=\utilde\@\cos\vphi-\vtilde\@\sin\vphi        \\
& \vtilde'=\utilde\@\sin\vphi+\vtilde\@\cos\vphi
\end{aligned}
\right.
\end{equation*}
The resulting direction $\@\psi\/(\Sigma)\subset\N_\a\@$ is similarly generated by a unit vector $\@\ztilde=\psi\/(\utilde)\@$, defined up to a phase factor
$\@\ztilde'=e^{\@i\@\vphi}\ztilde\@$.
\end{itemize}
\end{remark}

\begin{remark}\label{Rem2.2}
The $\@k^{\@th}\@$ normal harmonic involved in the representation \eqref{2.10} of the solutions of the linearized equations of motion coincides with the traditional one
if and only if the vectors $\@\htilde_{\@k}$ and $\@\rtilde_{\@k}$ are \emph{parallel\/}, i.e.~if and only if the vector $\ztilde_{\@k}=\htilde_{\@k}+i\rtilde_{\@k}\@$
is (proportional to) a real one.

In view of eq.~\eqref{2.9}, this happens if and only if the system
\begin{equation*}
\left\{
\begin{aligned}
&(C-\w^2\@A)\ztilde\@=\@0     \\[-2pt]
&\,B \ztilde\@=\@0
\end{aligned}
\right.
\end{equation*}
admits non--trivial solutions.

The simplest instance occurs in the case $\@B=0\@$, corresponding to the ordinary theory of small oscillations.
\end{remark}

\end{document}